\documentclass[12pt]{iopart}

\usepackage{epsfig}  
\begin{document}

\title[$H^0 \to \mu^+ \mu^-$ in multi-TeV $e^+e^-$ Collisions]{Testing the 
Higgs Mechanism in the Lepton Sector with multi-TeV $e^+e^-$ Collisions}

\author{M Battaglia}

\address{Department of Physics, University of California at Berkeley and \\ 
Lawrence Berkeley National Laboratory, Berkeley, CA 94720, USA}
\ead{MBattaglia@lbl.gov}

\begin{abstract}
Multi-TeV $e^+e^-$ collisions provide with a large enough sample of $H^0$ bosons 
to enable measurements of its suppressed decays. Results of a detailed study of 
the determination of the muon Yukawa coupling at $\sqrt{s}$ = 3~TeV, 
based on full detector simulation and event reconstruction, are presented. 
The muon Yukawa coupling can be determined with a relative accuracy of 0.04 to 
0.08 for Higgs bosons masses from 120~GeV to 150~GeV, with an integrated 
luminosity of 5~ab$^{-1}$. The result is not affected by overlapping
$\gamma \gamma \to {\mathrm{hadrons}}$ background.
\end{abstract}

\submitto{\JPG}
%\maketitle

\section{Introduction}

The detailed investigation of the Higgs sector is anticipated as one of the central themes 
of the accelerator particle physics program after the Higgs boson will have been observed, 
most likely at the LHC. The study of the Higgs profile will have to determine whether the 
observed particle is indeed responsible for generating the mass of gauge bosons, quarks and 
charged leptons. 
A significant {\it corpus} of studies, outlining strategies to test the Higgs mechanism for 
gauge bosons and quarks with accuracies down to a few percent, has already been 
assembled~\cite{Battaglia:2000jb, Barklow:2004th}
These rely on the combination of data from the LHC and an $e^+e^-$ linear collider 
operating at centre-of-mass energies $\sqrt{s}$ = 0.25~TeV - 1.0~TeV. 
In this paper we discuss the feasibility of a test of the Higgs mechanism in the lepton 
sector, by verifying the scaling $\frac{g_{H \mu \mu}}{g_{H \tau \tau}}$, through a precise 
determination of the $g_{H \mu \mu}$ coupling in the $H^0 \to \mu^+\mu^-$decay, in 3~TeV 
$e^+e^-$ collisions. 
At the LHC, the significance of an observation of the $H^0 \to \mu^+\mu^-$ decay is estimated 
to be only $\simeq$~3~$\sigma$ for 115~GeV $< M_H <$ 130~GeV, even when combining the CMS and 
ATLAS data on both the gluon and weak boson fusion production channels for 300~fb$^{-1}$ of 
integrated luminosity~\cite{Han:2002gp}. It further decreases for larger values of $M_H$.
$e^+e^-$ collisions below 1.0~TeV should provide with determination of $g_{H \tau \tau}$ to a 
relative statistical accuracy of $\simeq$~0.035 for a 120~GeV Higgs boson but, at best, 
with only a signal and not with a precision measurement in the $g_{H\mu\mu}$ Yukawa 
coupling~\cite{Battaglia:2001vf}. The $WW$ fusion process, $e^+e^- \to H^0 \nu_e \bar \nu_e$, 
offers a large Higgs cross section, provided high enough $\sqrt{s}$ energies can be attained, 
to make this measurement possible. In this paper the detection of the $H^0 \to \mu^+ \mu^-$ 
decay in high energy $e^+e^-$ collisions is studied using full {\tt Geant-4}-based simulation 
and reconstruction. The paper is organised as follows. 
In section~2 we discuss the simulation of the $e^+e^- \to H^0 \nu_e \bar \nu_e$, 
$H^0 \to \mu^+ \mu^-$ process and of the relevant physics and machine-induced background. 
Section~3 describes the detector simulation and section~4 has the results.

\section{$e^+e^- \to H^0 \nu_e \bar \nu_e$, $H^0 \to \mu^+ \mu^-$ at $\sqrt{s}$ = 3 TeV}

Higgs boson production in $e^+e^-$ collisions proceeds through two main processes: the 
so-called Higgstrahlung, $e^+e^- \to H^0 Z^0$, and the $WW$ fusion, 
$e^+e^- \to H^0 \nu_e \bar \nu_e$ reactions. 
While the associated production of the Higgs boson with a $Z^0$ has a number of advantages, 
including the possibility to tag the presence of the Higgs boson independent on its decay modes, 
$WW$ fusion reaches the largest cross sections, due to its $\propto \log \frac{s}{M^2_H}$ rise, 
if large enough centre-of-mass energies can be obtained. This makes the $WW$ fusion process of 
special interest for the study of rare Higgs decays, such as $H^0 \to b \bar b$ at large 
Higgs boson masses, and $H^0 \to \mu^+ \mu^-$. But as the energy increases, the $H^0$ production 
cross section is more and more peaked along the beam axis. Above $\simeq$~3~TeV, the further 
increase in production cross section is effectively lost, due to the limited detector acceptance 
in the very forward region. The production cross section at 3~TeV and 5~TeV is respectively 
2.3 and 3.0 times larger compared to that at 1~TeV. However, restricting the acceptance to 
Higgs bosons produced at $| \cos \theta | <$ 0.99 (0.92), the gains in cross section become 
2.1 (1.7) and 2.4 (1.9), respectively.  
The possibility of achieving $e^+e^-$ collisions at multi-TeV energies, introduced by the 
CLIC concept~\cite{Assmann:2000hg}, 
offers a real opportunity for observing this decay and measuring  $g_{H\mu\mu}$ with 
enough accuracy to perform a significant test of the scaling of the Yukawa couplings in the 
lepton sector over a broad range of Higgs boson masses~\cite{Battaglia:2001vf}. 
Multi-TeV $e^+e^-$ collisions become important not only for exploring the highest 
energy frontier but also for precision studies of suppressed Higgs decays.
 
\section{Data Analysis}

Signal events $e^+ e^- \to H^0 \nu_e \bar \nu_e$, $H^0 \to \mu^+ \mu^-$ are generated 
with {\tt Pythia 6.205}~\cite{Sjostrand:2000wi}, including beamsstrahlung effects, 
at several Higgs boson mass values from 120~GeV up to 160~GeV. 
Cross sections are computed using {\tt CompHep 4.4.0}~\cite{Boos:2004kh} and the Higgs 
decay branching fractions with {\tt HDECAY}~\cite{Djouadi:1997yw}.
At $\sqrt{s}$ = 3~TeV, the effective $e^+e^- \to H^0 \nu_e \bar \nu_e$ production cross 
section, accounting for initial state radiation, is 0.48~pb to 0.45~pb and the decay 
branching fraction, BR($H^0 \to \mu^+ \mu^-$) = 2.56$\times$10$^{-4}$ to 6.47$\times$10$^{-5}$,  
for $M_H$ in the range from 120~GeV to 150~GeV. 
The irreducible $\mu^+ \mu^- \nu_{\ell} \bar \nu_{\ell}$ background is generated 
with {\tt CompHep 4.4.0} interfaced with {\tt Pythia 6.205}. 
The total cross section of events with an invariant mass of the $\mu^+ \mu^-$ system, 
$M_{\mu\mu}$, in the range 100~GeV $< M_{\mu\mu} <$ 400~GeV is 5.31~fb. 
We assume to operate CLIC at centre of mass energy of 3~TeV for a total integrated 
luminosity of 5~ab$^{-1}$, which corresponds to $\simeq$~eight years (1 yr = 10$^7$~s) 
of operation at its nominal luminosity of 6.5$\times$10$^{34}$~cm$^{-2}$~s$^{-1}$. 
The CLIC anticipated parameters are given in~\cite{clic-params}, for this analysis we 
use the luminosity spectrum as in~\cite{Battaglia:2004mw}. 
This provides with $\simeq$~2.3$\times$10$^6$ $e^+e^- \to H^0 \nu_e \bar \nu_e$ events, 
620 to 150 of which are followed by the $H^0 \to \mu^+ \mu^-$ decay, depending on 
the value of the Higgs bosons mass in the range from 120~GeV to 150~GeV.  The machine 
induced background, which affects this analysis most, is due to parasitic 
$\gamma \gamma \to {\mathrm{hadrons}}$ events occurring within the same time-stamp as a 
physics event. The anticipated average number of such events for the CLIC parameters is 
3.2~BX$^{-1}$, or 161 events within a 25~ns time-stamp. $\gamma \gamma$ collisions are 
simulated using the {\tt GUINEAPIG} program~\cite{guineapig1, guineapig2} and the cross 
sections parametrised according to~\cite{Schuler:1996en}. 
The hadronic final states are simulated using {\tt Pythia} and overlayed to a 
signal event. The average visible energy deposited in the detector by 
$\gamma \gamma \to {\mathrm{hadrons}}$ events is $\simeq$~90~GeV/BX.

Charged particles are reconstructed in a multi-layered main tracker, made of high 
resolution silicon strip detectors, and a precision vertex tracker in a 5~T solenoidal 
field. Discrete Si tracking, following the example of the CMS experiment at LHC, emerged 
as an appealing solution for CLIC, where local hit occupancy and two-track separation may 
limit the applicability of gaseous trackers~\cite{Frey:2000eb,Battaglia:2004mw}. 
A large magnetic field is beneficial for ensuring an excellent momentum resolution for 
high $p_t$ particle tracks, as well as for confining low momentum charged particles from 
machine-induced backgrounds to small radii.
This analysis is based on an implementation of the SiD detector~\cite{sid}, developed as a 
concept for the International Linear Collider (ILC) project, for the reconstruction of 
3~TeV $e^+e^-$ events. 
The SiD concept has a five-layered Si main tracker which ensures a momentum resolution 
$\delta p_t/p_t^2 <$ 5$\times$10$^{-5}$ and polar angle coverage down to 8.3$~{\circ}$. 
The following assumptions on the detector response 
are made: the detection efficiency is 0.98 per layer, the point resolution 7~$\mu$m, which 
includes alignment effects, and the detector is capable of time-stamping hits with 25~ns 
resolution, corresponding to 50 bunch crossings (BX).
 
While the SiD is designed for the reconstruction of events at centre-of-mass energies 
below 1~TeV, its geometry is well suited for the study of simple events, as
those of our signal, up to multi-TeV energies.
Full event simulation is performed using the SiD model implemented in the 
{\tt Mokka 06-03} program~\cite{Musat:2004sp} based on {\tt Geant-4}~\cite{Agostinelli:2002hh}. 
Simulation results are saved in the {\tt lcio} format~\cite{Gaede:2003ip} and used as 
input to the event reconstruction. This is performed using a set of dedicated processors 
implemented in the {\tt Marlin} reconstruction and analysis framework~\cite{Gaede:2006pj}. 
The reconstruction starts from hits generated in the sensitive layers along the track. 
The track pattern recognition is performed from the outermost main tracker hits and 
extends inwards to the vertex tracker. After the first stage of pattern recognition
hit bundles are fitted using a simple helix. A second stage pattern recognition is  
performed using hits left unassociated after the first stage. Reconstructed particle 
tracks are requested to have more than 4 hits and associated hits on at least 
70~\% of the number of sensitive surfaces traversed, depending on their polar angle, 
$\theta$, $\chi^2$ fit probability in excess of 0.01, $| \cos \theta| <$ 0.98, and 
impact parameter significance below 3.5~times the extrapolation resolution, in both 
coordinates. These cuts remove poorly measured particle tracks. 
The momentum resolution of the reconstructed muon tracks is  
$\delta p_t/p_t^2$ = (3.46 $\pm$ 0.10)$\times$10$^{-5}$, in agreement with the 
SiD specifications. Events with two, oppositely charged muon candidate tracks with 
momentum exceeding 10~GeV and below 750~GeV and event energy in charged particles 
100~GeV $< E_{\mathrm{charged}} <$ 1.5~TeV are selected. 

\section{Results}

\begin{figure}[h!]
\begin{center}
\epsfig{file=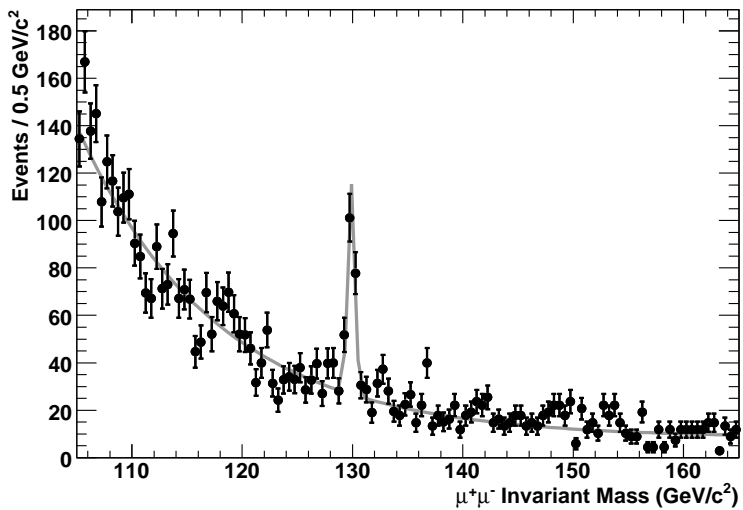,width=11.0cm}
\end{center}
\caption{\label{fig:mass}Reconstructed invariant mass distribution 
for signal and background $\mu^+ \mu^-$ + $E_{\mathrm{missing}}$ 
events with $M_H$ =130~GeV and 5~ab$^{-1}$ of integrated luminosity.}
\end{figure}

\begin{figure}[h!]
\begin{center}
\epsfig{file=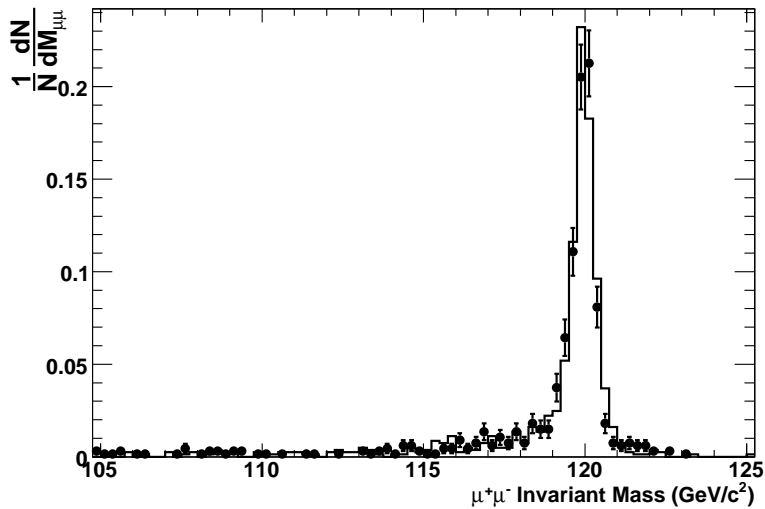,width=11.0cm}
\end{center}
\caption{\label{fig:comp}Reconstructed invariant mass distribution 
for signal events with $M_H$ =120~GeV without (histogram) and with 
(points with error bars) $\gamma \gamma \to {\mathrm{hadrons}}$ background 
overlayed.}
\end{figure}

The invariant mass distribution of the selected di-muon pairs is shown in Figure~\ref{fig:mass}. 
The distribution is fitted with a Gaussian curve to describe the signal and an exponential 
background. 
The mass resolution $\sigma_{M_H}/M_H$ for signal events is 0.0034, which corresponds 
to 0.40~GeV for $M_H$=120~GeV. The signal region is defined by a $\pm$~2~$\sigma_{M_H}$ interval 
around the fitted mass peak position and a binned $\chi^2$ fit is performed to extract the number 
of signal and background events. Results are summarised in 
Table~\ref{tab:results}.
\begin{table}
\caption{\label{tab:results}Number of selected signal and background events.}
\begin{indented}
\item[]\begin{tabular}{@{}lcccc}
\br
$M_H$ (GeV) & Nb.\ Signal Evts.\ & Nb.\ Bkg.\ Evts.\ & S/$\sqrt{\mathrm{B}}$ & 
$\delta{\mathrm{BR}}/{\mathrm{BR}}$ \\
\mr
120 & 229.6 & 161.1 & 18.1 & 0.086 \\
130 & 153.1 & ~88.1 & 16.3 & 0.101 \\
140 & 103.2 & ~64.3 & 12.9 & 0.125 \\
150 & ~68.1 & ~58.1 & ~9.5 & 0.160 \\
155 & ~68.1 & ~58.0 & ~5.2 & 0.253 \\
160 & ~12.1 & ~33.0 & ~2.1 & \\
\br
\end{tabular}
\end{indented}
\end{table}
We obtain
$\frac{\delta {\mathrm{BR(}}H^0 \to \mu\mu {\mathrm{)}}}
{{\mathrm{BR(}}H^0 \to \mu\mu {\mathrm{)}}}$ = 0.086 and 0.160 for $M_H$ = 120~GeV and 150~GeV, 
respectively. These results correspond to a determination of $g_{H \mu \mu}$ with a relative 
accuracy of 0.04 to 0.08. A significant signal for the decay can be obtained up to $M_H$ = 155~GeV, 
when its branching fraction is only 4$\times$10$^{-5}$, due to the rapid turn on of the $WW$ 
decay channel of the Higgs boson.
The analysis is repeated by overlaying the $\gamma \gamma \to {\mathrm{hadrons}}$ background 
on signal events, for $M_H$ = 120~GeV/$c^2$. No degradation of the reconstruction efficiency 
and resolution is observed for the amount of hadronic background corresponding to 50~BX 
(see Figure~\ref{fig:comp}).

\section{Conclusions}

The determination of the muon Yukawa coupling $g_{H \mu \mu}$ in $e^+e^-$ collisions at 
centre-of-mass energies of 3~TeV at CLIC has been studied through the 
$e^+e^- \to H^0 \nu_e \bar \nu_e$, $H^0 \to \mu^+ \mu^-$ process using full simulation 
and event reconstruction. The process is observable for Higgs mass values up to 155~GeV 
and the $g_{H \mu \mu}$ Yukawa coupling can be determined with a relative statistical 
accuracy of 0.04 to 0.08 for masses from 120~GeV to 150~GeV with an integrated 
luminosity of 5~ab$^{-1}$. The superposition of machine-induced 
$\gamma \gamma \to {\mathrm{hadrons}}$ background corresponding to 50 bunch crossings 
overlayed on a single $\mu^+ \mu^- \nu \bar \nu$ event does not affect these results.

\end{document}